%% file: paper.tex
\documentclass[a4paper,english,12pt,preprint,showpacs]{revtex4}
\usepackage[T1]{fontenc}
\usepackage[latin9]{inputenc}
\usepackage{verbatim}
\usepackage{graphicx}
\usepackage{amssymb}
\usepackage{esint}

\makeatletter

\usepackage{babel}
\makeatother

\begin{document}

\title{Mixing-induced CP violating sources for electroweak baryogenesis
from a semiclassical approach}

\author{Yu-Feng Zhou}

\affiliation{Korea Institute for Advanced Study, Seoul 130-722, Korea}

\email{yfzhou@kias.re.kr}

\begin{abstract}
The effects of flavor mixing in electroweak baryogenesis is investigated
in a generalized semiclassical WKB approach. Through calculating the
nonadiabatic corrections to the particle currents it is shown that
extra CP violation sources arise from the off-diagonal part of the
equation of motion of particles moving inside the bubble wall. This
type of mixing-induced source is of the first order in derivative
expansion of the Higgs condensate, but is oscillation suppressed.
The numerical importance of the mixing-induced source is discussed
in the Minimal Supersymmetric Standard Model and compared with the
source term induced by semiclassical force. It is found that in a
large parameter space where oscillation suppression is not strong
enough, the mixing-induced source can dominate over that from the
semiclassical force. 
\end{abstract}


\preprint{KIAS-TH-P08037}
\preprint{NSF-KITP-08-73}

\maketitle

\section{Introduction}

Electroweak baryogenesis (EWBG) is a promising scenario for explaining
baryon number asymmetry in the universe, unlike other scenarios valid
at grand unification scale it can be tested by the upcoming collider
experiments. The Standard Model (SM) of particle physics due to the
too small CP violation \cite{Farrar1994,Farrar1993,Gavela1994,Huet1995}
and difficulty in generating a strongly first order phase transition
\cite{Jansen1996,Rummukainen1998a,Rummukainen1998} can not be a
candidate for baryogenesis. However, EWBG is viable in many new physics
models beyond the SM such as Minimal Supersymmetric Standard Model
(MSSM) \cite{Espinosa1993,Giudice1992,Myint1992} and the two-Higgs-doublet
models \cite{Cline:1995dg,Cline:1996mga,Funakubo:1996dw,Funakubo:1997bm,Hammerschmitt:1994fn,Kanemura:2004ch,Land:1992sm,Turok:1990zg}
etc, as in these models there are extra particles contributing to
the Higgs potential and also new sources of CP violation.

An efficient way to generate large baryon number asymmetry is to generate
it nonlocally, through a charge transportation mechanism \cite{Cohen1991,Cohen:1990py,Nelson:1991ab}
in which the CP violating charges are first generated inside the bubble
wall and then get transported into the unbroken phases, where the
anomalous baryon number violating sphaleron processes are not suppressed.
In some models such as MSSM and the two-Higgs-doublet model, the bubble
wall of the electroweak phase transition is typically thick \cite{Cline1998,Moreno1998}
for the particles giving dominant contributions, and the particle
typical Compton wave length $\lambda\sim1/T$ is much shorter than
the wall width $L_{w}\sim(10-20)/T$ where $T$ is the critical temperature.
In this regime the validity of semiclassical approach \cite{Cline:2001rk,Cline:2000nw,Cline:1997vk,Cline2000,Huber:2000mg,Joyce1996,Joyce1995,Joyce1994}
should be justified for single flavor case, which provides an intuitively
simple description by treating particle transportation as motion of
WKB wave packages. In a slowly moving and CP violating Higgs condensate
background, the dispersion relation for particles and antiparticles
are modified differently, contributing to different semiclassical
forces, which leads to a net excess or deficit of the particle number
which can be converted into left-handed fermion number asymmetry.
The asymmetry of the local fermion density then get transported in
front of the bubble wall, which bias the baryon number violating processes. 

The problem becomes more involved in the multiple flavor case, where
the CP violating mass matrix has nontrivial space-time dependence.
The CP violating effects may show up not only in the dispersion relations,
but also in the mixing of states. In the conventional WKB approach
\cite{Cline:2001rk,Cline:2000nw,Huber:2000mg} the particle propagation
is treated as an adiabatic motion: the quasiparticles are first rotated
into the local mass eigenstates, then these particles are assumed
to evolve individually without interference. The semiclassical force
terms are identified for each individual states as the unique CP violating
source for the diffusion equations. However, note that in many cases
such as in MSSM, the quasiparticles do not decouple even in the local
mass basis, because the rotation matrices diagonalizing the mass matrix
are also spatially varying, which leads to interferences among the
local mass eigenstates. Namely, there always exists nonvanishing off-diagonal
terms in the equations of motion. The off-diagonal terms in the equation
of motion lead to nonconservation of net currents, which contributes
to extra CP violating source terms. Such a contribution can be numerically
important, because the semiclassical force term is of the second order
in derivative expansion, and is further suppressed to the third order
when entering the semiclassical diffusion equations as a source term,
while the mixing-induced source presents from the first order. 

The problem of spatially dependent mixing has been discussed in the
non-equilibrium field theory approach based on the Kadanoff-Baym equations
for Wigner transformed Green functions \cite{Prokopec2004,Prokopec2004a}.
It has been noticed recently that the evolution of the off-diagonal
densities from the transverse part of the constraint equation exhibit
oscillation behavior, which is quite different from the diagonal densities
\cite{Konstandin2005,Konstandin2006}. Other methods although using
the same Schwinger-Keldysh formalism treat the problem in different
manner, such as using the Higgs condensate insertion \cite{Lee2005,Riotto1998,Riotto1998a}
or applying re-summation for Higgs condensate insertion but using
phenomenological ways to deal with the source terms \cite{Carena:2000id}.
The final result still differ significantly, especially in the bosonic
sector. 

In view of the current theoretical situation, a carefully reanalysis
of the semiclassical method may still be useful. In this work we generalize
the semiclassical method by taking into account the non-adiabatic
corrections from the spatially varying flavor mixings. We restrict
ourselves in the parameter region where the mixing between the local
mass eigenstates are relatively small, and can be treated as perturbations
to the equation of motion for local mass eigenstates. Our results
show that the correction leads to an extra CP violating source which
exhibit a typical oscillation behavior in analogy to neutrino oscillations,
which is similar to the observations made in \cite{Konstandin2005}.
We go a step further to estimate the source term and compare it with
the semiclassical forces in the context of MSSM. The results show
that despite the oscillation suppression, the mixing-induced source
term has significant numerical importance for a large parameter space.

\section{The fermionic case}

We begin with a brief review of the semiclassical description for
fermions in a expanding planar bubble wall. For simplicity the fermion
is boosted into a wall frame in which the momentum parallel to the
wall is vanishing. In this case the equation of motion is reduced
to 1+1 dimensional, and in flavor basis it is given by\begin{equation}
[i(\gamma_{0}\partial_{t}+\gamma_{3}\partial_{z})-M^{\dagger}(z)P_{L}-M(z)P_{R}]\psi=0\label{eq:EOM-flavor}\end{equation}
where $M(z)$ is the spatial varying mass matrix from a mass term
$\bar{\psi}_{L}M(z)\psi_{R}$ induced by a slowly moving Higgs condensate,
and $P_{L,R}=(1\mp\gamma_{5})/2$ the chirality projection operator.
The wave function in the chiral representation of Dirac matrices can
be decomposed in terms of spin eigenstates\begin{equation}
\psi_{s}=e^{-i\omega t}\left(\begin{array}{c}
L_{s}\\
R_{s}\end{array}\right)\otimes\chi_{s}\end{equation}
with $\sigma_{3}\chi_{s}=s\chi_{s}$ and $s=\pm1$ for two spin states%
{}. $L_{s}(R_{s})$ is the left- (right-) handed component. Substituting
$\psi_{s}$ into Eq.(\ref{eq:EOM-flavor}) and eliminating the right-(left-)
handed component one obtains the equation for $L_{s}(R_{s})$ in flavor
basis %
{}  \begin{eqnarray}
\left[\omega^{2}+\partial_{z}^{2}-MM^{\dagger}+isM\partial_{z}M^{-1}(\omega-is\partial_{z})\right]L_{s} & = & 0\\
\left[\omega^{2}+\partial_{z}^{2}-M^{\dagger}M-isM^{\dagger}\partial_{z}M^{\dagger-1}(\omega+is\partial_{z})\right]R_{s} & = & 0\label{eq:EOM-flavorBasis}\end{eqnarray}
One can rotate the fields into a local mass basis ($L_{s}^{d},R_{s}^{d}$)
by defining $L_{s}=UL_{s}^{d}$ and $R_{s}=VR_{s}^{d}$ in which the
mass matrix %
{} $M(z)$ is diagonalized $U^{\dagger}M(z)V=m(z)$ with $m(z)$ a diagonal
mass matrix. The phases of $U$ and $V$ are arranged such that the
mass eigenvalues $m_{i}$ are real positive. Note that both $U$ and
$V$ are $z$ dependent. The equation of motion for local mass eigenstates
$L_{s}^{d}$ and %
{}and $R_{s}^{d}$ becomes\begin{eqnarray}
i(\partial_{t}-s\partial_{z}-sU^{\dagger}\partial_{z}U)L_{s}^{d}-mR_{s}^{d} & = & 0\\
i(\partial_{t}+s\partial_{z}+sV^{\dagger}\partial_{z}V)R_{s}^{d}-mL_{s}^{d} & = & 0\end{eqnarray}
It is evident that the equation for right-handed component in the
local mass basis can be obtained by replacing $U\to V$ and $s\to-s$.
Following the similar step in obtaining Eq.(\ref{eq:EOM-flavorBasis})
one arrives at the decoupled second order equations %
{}\begin{eqnarray}
\left[\omega^{2}+\partial_{z}^{2}-m^{2}+2\Sigma\partial_{z}+isA(\omega-is\partial_{z})\right]L_{s}^{d} & = & 0\\
\left[\omega^{2}+\partial_{z}^{2}-m^{2}+2\Pi\partial_{z}-isB(\omega+is\partial_{z})\right]R_{s}^{d} & = & 0\end{eqnarray}
where \begin{eqnarray}
\Sigma=U^{\dagger}\partial_{z}U & , & A=U^{\dagger}M\partial M^{-1}U\\
\Pi=V^{\dagger}\partial_{z}V & , & B=V^{\dagger}M^{\dagger}\partial M^{\dagger-1}V\end{eqnarray}
%
{} In the above equations the terms of second order in derivative expansion
are neglected. The off-diagonal elements of $A$ and $B$ are related
to $\Sigma$ and $\Pi$ through\begin{eqnarray*}
A_{ij}=\frac{m_{i}}{m_{j}}\Pi_{ij}-\Sigma_{ij} & , & B_{ij}=\frac{m_{i}}{m_{j}}\Sigma_{ij}-\Pi_{ij}\ \mbox{ ( for }i\neq j)\end{eqnarray*}
In the remainder of this section we shall suppress the index {}``$d$''
in the wave function.
In two flavor mixing case the explicit form of the equation array
for left-handed component can be rewritten as

\begin{equation}
\left(\begin{array}{cc}
D_{11} & D_{12}\\
D_{21} & D_{22}\end{array}\right)\left(\begin{array}{c}
L_{s1}\\
L_{s2}\end{array}\right)=0\label{eq:EOM-mass}\end{equation}
with\begin{eqnarray}
D_{11} & = & \omega^{2}+\partial_{z}^{2}-m_{1}^{2}+2\Sigma_{11}\partial_{z}+isA_{11}(\omega-is\partial_{z})\\
D_{12} & = & 2\Sigma_{12}\partial_{z}+isA_{12}(\omega-is\partial_{z})\\
D_{21} & = & 2\Sigma_{21}\partial_{z}+isA_{21}(\omega-is\partial_{z})\\
D_{22} & = & \omega^{2}+\partial_{z}^{2}-m_{2}^{2}+2\Sigma_{22}\partial_{z}+isA_{22}(\omega-is\partial_{z})\end{eqnarray}
Since the off-diagonal elements $D_{12(21)}$ contain differential
operators, one can not obtain decoupled equations for $L_{s1(2)}$
separately. In the conventional approach \cite{Cline:2000nw}, the
equation array is solved approximately by first imposing the derivative
operator $D_{22(11)}$ on the first (second) row of Eq.(\ref{eq:EOM-mass})
once again, then picking up the terms up to the first order derivative,
which leads to equations containing higher order derivatives $D_{ii}D_{jj}L_{sj}\simeq0(i\neq j)$
and finally argue that they reduce to a decoupled form\begin{equation}
D_{ii}L_{si}=0,\label{eq:EOM-diagonal}\end{equation}
which is essentially the diagonal part of Eq.(\ref{eq:EOM-mass}).
Note that in this method the off diagonal terms are counted as higher
order terms and thus neglected. An advantage of this method is that
it naturally allows an adiabatic description of the motion of individual
particles. However, It is clear that Eq.(\ref{eq:EOM-diagonal}) misses
the information from off-diagonal term $D_{12,21}$, which is only
valid when $\left|D_{12,21}\right|\ll\left|D_{11}-D_{22}\right|$.
In large mixing case it is not a good approximation to Eq.(\ref{eq:EOM-mass}). 

To investigate the impact of the off-diagonal terms, in this work
we adopt an alternative method to solve Eq.(\ref{eq:EOM-mass}) approximately.
We are interested in the parameter region in which the off-diagonal
elements is non-negligible but relatively small such that it can be
treated as perturbations. The validity of the perturbation requires
that the mass difference between the two mass eigenstates are significantly
larger than the mixing term, namely $\partial_{z}m\ll\left|m_{i}^{2}-m_{j}^{2}\right|$,
where $m$ is the averaged mass. Although this method does not apply
to the resonance case $\left|m_{i}^{2}-m_{j}^{2}\right|\simeq0$,
it should illustrate the main feature of the mixing caused by off-diagonal
term $D_{ij}$. 

Taking the off-diagonal terms $D_{12}$ and $D_{21}$ as perturbations.
The solutions can be written in a generic form \begin{equation}
L_{si}=L_{si}^{(0)}+L_{si}^{(1)}\end{equation}
where $L_{si}^{(0)}$ is the lowest order solution satisfying Eq.(\ref{eq:EOM-diagonal}),
i.e. \begin{equation}
D_{ii}L_{i}^{(0)}=0\label{eq:LowestOrderEq}\end{equation}
and $L_{si}^{(1)}$ are the corrections due to the off-diagonal terms.
The lowest order solution $L_{si}^{(0)}$ is obtained by the usual
WKB wave ansatz

\[
L_{si}^{(0)}=w_{i}e^{i\int^{z}p_{ci}(z')dz'}\]
where $p_{ci}$ is the canonical momentum and the function $w_{i}$
provides the correct normalization for $L_{si}^{(0)}$. The real and
imaginary part of Eq.(\ref{eq:LowestOrderEq}) lead to two separated
equations%
{}\begin{eqnarray}
\omega^{2}-p_{ci}^{2}-m_{i}^{2} & = & \mbox{Im}\left[2p_{ci}\Sigma_{ii}+s(\omega+sp_{ci})A_{ii}\right]\\
p_{ci}'+2p_{ci}\frac{w'_{i}}{w_{i}} & = & \mbox{Re}\left[2p_{ci}\Sigma_{ii}+s(\omega+sp_{ci})A_{ii}\right]\end{eqnarray}
%
{}where the notation prime stands for the spatial derivative $\partial_{z}$.
The first equation gives the dispersion relation\begin{equation}
p_{ci}\simeq p_{0}-\frac{s(\omega+sp_{0})}{2p_{0}}\mbox{Im}A_{ii}-\mbox{Im}\Sigma_{ii}+\alpha'\end{equation}
where $\alpha$ is an arbitrary phase factor from gauge invariance
\cite{Cline2000}. From the dispersion relation one can deduce the
group velocity $v_{g}\equiv(\partial\omega/\partial p_{c})_{z}$ and
semiclassical force $F_{i}\equiv\omega(dv_{g}/dt)$\begin{eqnarray}
v_{gi} & = & \frac{p_{0i}}{\omega}-s\frac{m_{i}^{2}\mbox{Im}A_{ii}}{2\omega^{2}p_{0i}}\\
F_{i} & = & -\frac{m_{i}m_{i}'}{\omega}-s\frac{\left(m_{i}^{2}\mbox{Im}A_{ii}\right)'}{2\omega^{2}}\label{eq:force}\end{eqnarray}
 Note that only the second term in the force term is CP violating,
and is proportional to the spin of the local mass states. Together
with the lowest order solution to $w_{i}$ from the second equation,
the wave function at the lowest order is given by%
{}\begin{eqnarray}
L_{i}^{(0)} & = & \frac{m_{i}}{\sqrt{p_{0i}(\omega+sp_{0i})}}e^{i\int^{z}p_{ci}dz'},\mbox{ and }R_{i}^{(0)}=\frac{m_{i}}{\sqrt{p_{0i}(\omega-sp_{0i})}}e^{i\int^{z}p_{ci}dz'}\end{eqnarray}
with $p_{0i}^{2}=\omega^{2}-m_{i}^{2}$%
{}. 

Substituting the off-diagonal terms in to the equation, the first
order perturbation takes the following form\begin{eqnarray}
L_{si} & \simeq & L_{si}^{(0)}+L_{si}^{(1)}=L_{si}^{(0)}+\epsilon_{i}L_{si}^{(0)}\\
R_{si} & \simeq & R_{si}^{(0)}+R_{si}^{(1)}=R_{si}^{(0)}+\delta_{i}L_{si}^{(0)}\end{eqnarray}
which are mixtures of the two unperturbed states. The mixing parameters
for particle $i$ to the first order of derivative are given by

\begin{equation}
\epsilon_{i}=i\frac{2\Sigma_{ij}p_{0j}+A_{ij}(s\omega+p_{0j})}{m_{i}^{2}-m_{j}^{2}},\mbox{ and }\delta_{i}=i\frac{2\Pi_{ij}p_{0j}+B_{ij}(-s\omega+p_{0j})}{m_{i}^{2}-m_{j}^{2}}\end{equation}
The mixing parameters for particle $j$ can be simply obtained by
replacing $i\leftrightarrow j$ from the above expressions. It is
clear that the expansion is valid for $\partial_{z}m_{i}/(\Delta m^{2})\ll1$
as expected. The momentum dependencies comes from the differentiation
operators in off-diagonal element $D_{12}$. 

During the particle propagation inside the bubble wall, the Higgs
condensate background causes mixing between the two local mass states,
which keeps the total particle current conserved but leads to nonconservation
of the currents for individual particles. The nonvanishing divergence of the current
caused by the off-diagonal part of spatially varying rotation matrix
is proportional to the mixing parameters $\epsilon_{i}$ and $\delta_{i}$,
which are different for particle and antiparticle when the mixing
contains CP violation. This provides an extra CP violating source
other than the semiclassical force. Note that the mixing term modifies
the dispersion relation as well. As it contains derivative terms,
it implies the breaking down of the semiclassical picture. However,
this is a next to leading order correction to the dispersion relation
and is expected to be small. 

As an illustration, we calculate the left-handed current divergence.
The temporal and spatial part of the left-handed current are $j_{}^{0}=L_{s}^{*}L_{s}\mbox{ and }j^{z}=-sL_{s}^{*}L_{s}$
respectively. The mixing parameters for the antiparticle can be obtained
by replacing $\Sigma\to\Sigma^{*},A\to A^{*},\Pi\to\Pi^{*},$ and
$B\to B^{*}$. The mixing-induced left-handed current, after subtracting
the antiparticle component is given by

\begin{equation}
j_{Li}^{z}=-\frac{4m_{i}m_{j}}{m_{i}^{2}-m_{j}^{2}}\frac{2s\mbox{Im}\Sigma_{ij}p_{0j}+\mbox{Im}A_{ij}(\omega+sp_{0j})}{\sqrt{p_{0i}p_{0j}(\omega+sp_{0i})(\omega+sp_{0j})}}\cos\int^{z}(p_{cj}-p_{ci})dz'\end{equation}

The CP violating force term, as it is proportional to spin $s$, only
contribute to the spin-weighted density. To facilitate the comparison
with the force term, we calculate the spin-weighted mixing-induced
source term. Indeed, it has been pointed out that only spin-weighted
current is sourced by the moving wall \cite{Konstandin2005}. In
the wall frame, the corresponding source term for the left-handed
current is \begin{eqnarray}
S_{Li} & = & \sum_{s}\frac{s}{2}\partial_{\mu}j_{Li}^{s\mu}\\
 & = & \frac{2m_{i}m_{j}}{m_{i}^{2}-m_{j}^{2}}\left[\mbox{Im}\Sigma g_{L}(p_{0i},p_{0j})-\frac{m_{i}}{m_{j}}\mbox{Im}\Pi_{ij}g_{R}(p_{0i},p_{0j})\right]\\
 &  & \cdot(p_{0i}-p_{0j})\sin\int^{z}(p_{cj}-p_{ci})dz'\label{eq:source}\end{eqnarray}
with\begin{equation}
g_{L,R}(p_{0i},p_{0j})=\frac{\omega\mp p_{0j}}{\sqrt{p_{0i}p_{0j}(\omega+p_{0i})(\omega+p_{0j})}}-\frac{\omega\pm p_{02}}{\sqrt{p_{0i}p_{0j}(\omega-p_{0i})(\omega-p_{0j})}}\end{equation}
which is a momentum odd function.

The numerical importance of the source term comes from the fact that
it is of the first order in derivative expansion, larger than that
from semiclassical force which is of the second order. Furthermore,
The final form of the semiclassical force term entering the semiclassical
diffusion equation is of the third order in derivative as it is weighted
by group velocity. While in the diffusion equation, the mixing-induced
source term is not suppressed. Therefore in a naive counting, one
expects a much larger contribution from mixing induced source term. 

What might significantly suppress the mixing-induced source is the
oscillation. A crucial feature is that the oscillation frequency depends
on the momentum {\em difference} rather than the momentum themselves,
which makes the oscillation suppression less effective for relatively
small momentum differences. From Eq.(\ref{eq:source}), the oscillation
term can be approximated by $\sin(p_{02}-p_{01})dz$ at the lowest
order. In a highly relativistic limit the oscillation wave length
$\lambda_{osc}$ is given by \[
\frac{1}{\lambda_{osc}}\simeq\frac{m_{1}^{2}-m_{2}^{2}}{4\pi\omega}\]
which is well-known for neutrino oscillations. For an illustration,
taking $\omega\simeq300$ GeV, $m_{1}\simeq200$GeV and $m_{2}\simeq100$
GeV, we find $\lambda_{osc}$ is around $\mathcal{O}(0.1\mbox{GeV}^{-1})$
which is much larger than the typical Compton wave length $1/T$.
Thus the oscillation is unlikely to completely erase the mixing-induced
source. 

Finally, The source is proportional to the off-diagonal elements of
$\Sigma_{ij}$ and $\Pi_{ij}$, which is different from the CP violating
source term which is always proportional to the diagonal elements
$A_{ii}$. This leads to a different parameter dependence in the final
source term.

\section{The Bosonic case}

The mixing in bosons can be calculated in a similar way. The Klein-Gordon
equation in flavor basis is\begin{equation}
\partial^{\mu}\partial_{\mu}\phi+M^{2}(z)\phi=0\end{equation}
where $M^{2}(z)$ is the mass square matrix which is diagonalized
by a unitary transformation $U^{\dagger}M^{2}(z)U=m^{2}(z)$. Rotating
$\phi$ into the local mass basis with $\phi=U\phi^{d}$, the equation
of motion for $\phi^{d}$ in the wall frame has the form \begin{equation}
\partial_{t}^{2}\phi^{d}-\partial_{z}^{2}\phi^{d}+2U^{\dagger}\partial^{z}U\partial_{z}\phi^{d}+m_{}^{2}\phi^{d}=0\end{equation}
We shall suppressed the index {}``$d$'' in the remainder of this
section. In the two flavor mixing case the above equation can be rewritten
in a matrix form

\begin{equation}
\left(\begin{array}{cc}
D_{11} & D_{12}\\
D_{21} & D_{22}\end{array}\right)\left(\begin{array}{c}
\phi_{1}\\
\phi_{2}\end{array}\right)=0\label{eq:EOM-boson}\end{equation}
The operators $D_{ij}$ are given by\begin{eqnarray}
D_{11} & = & \partial^{2}+m_{1}^{2}+2\Sigma_{11}\partial_{z}\\
D_{12} & = & 2\Sigma_{12}\partial_{z}\\
D_{21} & = & 2\Sigma_{21}\partial_{z}\\
D_{22} & = & \partial^{2}+m_{2}^{2}+2\Sigma_{22}\partial_{z}\end{eqnarray}
with $\Sigma\equiv U^{\dagger}\partial_{z}U$. Again we solve the
equation by taking off-diagonal elements $D_{12(21)}$ as perturbations.
The solution takes the form\begin{equation}
\phi_{i}=\phi_{i}^{(0)}+\phi_{i}^{(1)}\end{equation}
with $\phi_{i}^{(0)}$ satisfying the lowest order equations \begin{equation}
D_{ii}\phi_{i}^{(0)}=0\label{eq:diagonal}\end{equation}
Making use of the WKB ansatz \begin{equation}
\phi_{i}^{(0)}=w_{i}e^{-i\omega t}e^{i\int^{z}p_{ci}(z')dz'}\end{equation}
and substituting it into the Eq.(\ref{eq:diagonal}), the real and
imaginary part gives two independent equations\begin{eqnarray}
-\omega^{2}+p_{ci}^{2}+m_{i}^{2}-ip_{ci}'+2\mbox{Im}\Sigma_{ii}\cdot p_{ci} & = & 0\\
-p_{ci}'+2\frac{w'}{w}p_{ci}+2\mbox{Re}\Sigma_{ii}\cdot p_{ci} & = & 0\end{eqnarray}
The first equation indicates the modification of dispersion relation.
An important difference from the fermionic case is that the dispersion
relation for scalar fields will not generate the semiclassical force
at the first order in derivative expansion because the contribution
from $\mbox{Im}\Sigma$ cancels in the expression for $dv_{g}/dt$.
The second equation determines the form of $w$. To the lowest order
$w_{i}\simeq1/\sqrt{p_{ci}}$. Taking the mixing term $D_{ij}$ as
perturbations, the solution takes the form \begin{equation}
\phi_{i}^{}\simeq\phi_{i}^{(0)}+\epsilon_{i}\phi_{j}^{(0)},\end{equation}
the mixing coefficient to the first order in derivative is \begin{equation}
\epsilon_{i}=i\frac{2\Sigma_{ij}p_{jc}}{m_{i}^{2}-m_{j}^{2}}\end{equation}
The mixing parameter for particle $j$ can be obtained by replacing
$i\leftrightarrow j$ in the above expression. The spatial component
for the current is $j_{i}^{z}=i(\phi_{i}^{*}\partial^{z}\phi_{i}-\partial^{z}\phi_{i}^{*}\phi_{i})$.
Substitute the expression of $\phi_{i}$ the mixing-induced current
is given by\begin{equation}
j_{i}^{z}=i(\epsilon\phi_{i}^{(0)*}\partial\phi_{j}^{(0)}-\epsilon^{*}\partial\phi_{j}^{(0)*}\phi_{i}^{(0)})+i(\epsilon^{*}\phi_{j}^{(0)*}\partial\phi_{i}^{(0)}-\epsilon^{*}\partial\phi_{i}^{(0)*}\phi_{j}^{(0)})\end{equation}
The divergence of the current is%
{} \begin{equation}
S_{i}=\partial_{\mu}j_{i}^{\mu}=-2\mbox{Im}[\epsilon_{i}^{'}(\partial\phi_{i}^{(0)*}\phi_{j}^{(0)}+\partial\phi_{j}^{(0)*}\phi_{i}^{(0)})]\end{equation}
In obtaining the above expression, we have used the lowest order equation
of motion for $\phi_{i}^{(0)}$. Note that the terms proportional
to $\epsilon_{i}$ all canceled out. Only the $\epsilon'_{i}$ terms
remain. Therefore the source term is of second order in derivative,
which is of the next to leading order compared with the fermionic
case. This makes the contribution from boson mixing to be expected
less important.

\section{source terms in diffusion equation}

At the semiclassical level the divergence of the currents derived
in the previous section can be directly added to the Boltzmann equation
as semiclassical sources. In the wall frame the Boltzmann equation
has the form \begin{equation}
\frac{\partial f_{i}}{\partial t}+\frac{\partial f_{i}}{\partial z}v_{g}+\frac{\partial f_{i}}{\partial p_{z}}F_{iz}=C[f_{i},f_{j}\cdots]+S_{i}\end{equation}
where $f_{i}=f_{i}(t,x,p)$ is the phase space distribution function,
$C$ is the collision term and the $p_{z}$ the kinetic momentum defined
by $p_{z}\equiv\omega v_{g}$. In a fluid ansatz, the distribution
function $f_{i}$ in the wall frame is \begin{equation}
f_{i}=\frac{1}{e^{\beta[\gamma_{w}(\omega+v_{w}p_{z})-\mu_{i}(z)]}\pm1}+\delta f_{i}(z,p)\end{equation}
where the spatially varying chemical potential $\mu_{i}(z)$ describe
the departure from chemical, and the perturbation $\delta f_{i}(z,p)$
describe the response to the semiclassical force, which by definition
has no contribution to the particle density, i.e. $\int d^{3}p\delta f_{i}=0$.
The factor $v_{w}p_{z}$ comes from the Lorentz boost from the plasma
frame to the wall frame. Substituting the distribution function into
the Boltzmann equation \begin{equation}
\frac{\partial f_{i}}{\partial\omega}\left(v_{w}F_{iz}-\mu'\frac{p_{zi}}{\omega}\right)+\frac{p_{zi}}{\omega}\delta f'_{i}=C^{}[f_{i},f_{j}\cdots]+S_{i}\end{equation}
and boosting back into the plasma frame by a Galilean transformation
$v_{g}\to v_{g}+v_{w}$ which is valid for small $v_{w}\ll1$, we
have\begin{equation}
\left(\frac{p_{iz}}{\omega}+v_{w}\right)\left(-\mu'\frac{\partial f_{0i}}{\partial\omega}+\delta f_{i}^{'}\right)+v_{w}\frac{\partial f_{0i}}{\partial\omega}\delta F_{zi}=C_{i}^{pl}[\mu_{i},\delta f_{i}]+S_{i}^{pl}\end{equation}
with $f_{0i}$ the up-perturbed Fermi-Dirac distribution $f_{0}=(e^{\beta\omega}+1)^{-1}$
in the plasma frame. The quantity $\mu_{i}$ and $\delta F_{zi}$
contain only CP violating part of chemical potential and semiclassical
force. From Eq.(\ref{eq:force}), the CP violating force term is\begin{equation}
\delta F_{zi}=-s\frac{(m_{i}^{2}\mbox{Im}A_{ii})'}{2\omega^{2}}\end{equation}
The source term $S_{i}^{pl}$ in the plasma frame is obtained from
$S_{i}$ by a boost $p_{z}\to p_{z}+v_{w}\omega$, namely $S_{i}^{pl}=S_{i}(p_{z}+v_{w}\omega)$.
Note that the $p_{z}-$even part in $S_{i}^{pl}$ only comes from
$v_{w}\omega$ term which is nonvanishing after momentum integration.
Thus the source is proportional to the wall velocity $v_{w}$ and
is vanishing in the limit of $v_{w}\to0$, which is a physical requirement.
Integrating over $d^{3}p$ weighted by 1 and $p_{z}/\omega$ respectively,
we get two transportation equations\begin{eqnarray}
-v_{w}\frac{\mu_{i}^{'}}{T}+\left\langle \frac{p_{z}}{\omega}\delta f_{i}^{'}\right\rangle  & = & \left\langle C_{i}^{pl}\right\rangle +\left\langle S_{i}^{pl}\right\rangle \\
-\frac{\mu^{'}}{T}\left\langle \left(\frac{p_{z}}{\omega}\right)^{2}\right\rangle +v_{w}\left\langle \frac{p_{z}}{\omega}\delta f^{'}\right\rangle +\frac{v_{w}}{T}\left\langle \frac{p_{z}}{\omega}\delta F_{zi}\right\rangle  & = & \left\langle \frac{p_{z}}{\omega}C_{i}^{pl}\right\rangle +\left\langle \frac{p_{z}}{\omega}S_{i}^{pl}\right\rangle \label{eq:collison-term}\end{eqnarray}
where the definitions for the integration are \cite{Cline2000,Huber:2000mg}
\[
\left\langle X\right\rangle \equiv k_{i}\frac{\int d^{3}p\frac{\partial f_{0i}}{\partial\omega}X}{\int d^{3}p\frac{\partial f_{0i}}{\partial\omega}}(\mbox{for }X=\frac{p_{z}^{2}}{\omega^{2}},\frac{p_{z}}{\omega}\delta F_{zi}),\mbox{ and }\left\langle X\right\rangle \equiv k_{i}\frac{\int d^{3}pX}{T\int d^{3}p\frac{\partial f_{0i}}{\partial\omega}},(\mbox{ }X=\mbox{others)}\]
With the factor $k_{i}\equiv\int d^{3}p\frac{\partial f_{0i}}{\partial\omega}/\int d^{3}p\frac{\partial f_{0i}(m_{i}=0)}{\partial\omega}$
which is 1(2) for massless fermion(boson). The collision terms can
be expressed in terms of chemical potentials and inelastic interaction
rates\begin{eqnarray*}
\left\langle C_{i}^{pl}\right\rangle =-k_{i}\Gamma_{ik}^{d}\sum_{j}\xi_{j}^{(k)} & \mbox{ , } & \left\langle \frac{p_{z}}{\omega}C_{i}^{pl}\right\rangle \simeq k_{i}\left\langle (p_{z}/\omega)\delta f_{i}\right\rangle \Gamma_{i}^{t}\end{eqnarray*}
where $\Gamma_{ik}^{d}$ is rate for the inelastic channel $(k)$
and $\sum_{j}\xi_{j}^{(k)}$ are the signed sum over the relevant
chemical potential. The quantity $\Gamma_{i}^{t}$ is the total interaction
rate. Eliminate the quantity $\left\langle \frac{p_{z}}{\omega}\delta f_{i}^{'}\right\rangle $
by differentiating the second equation in Eq.(\ref{eq:collison-term})
once again, one arrives at the usual form of the diffusion equation
%
{}\[
-k_{i}\left(D_{i}\xi_{i}^{''}+v_{w}\xi'\right)+\tilde{\Gamma}_{ik}^{d}\sum_{j}\xi_{j}^{(k)}\simeq S_{F}+S_{M}\]
with $\tilde{\Gamma}_{ik}^{d}=k_{i}\Gamma_{ik}^{d}$\[
S_{F}=-\frac{k_{i}v_{w}D_{i}}{T\left\langle \left(\frac{p_{z}}{\omega}\right)^{2}\right\rangle }\left\langle \frac{p_{z}}{\omega}\delta F_{zi}\right\rangle ',\mbox{ and }S_{M}=k_{i}\left\langle S_{i}^{pl}\right\rangle \]
where $\xi\equiv\mu_{i}/T$ the rescaled chemical potential and $D_{i}=\left\langle (p_{z}/\omega)^{2}\right\rangle /\Gamma_{i}^{t}$
the diffusion constant. In deriving the diffusion we have neglected
the $\left\langle (p_{z}/\omega)S_{i}^{pl}\right\rangle '$ term which
is subleading compared with $\left\langle S_{i}^{pl}\right\rangle $.

\section{mixing induced-source in mssm}

In MSSM, the chargino transportation provides a dominant CP violating
source to electroweak baryogenesis. The asymmetry in chargino number
is converted into the asymmetry in left-handed top quarks through
Yukawa interactions. The two Higgsino $SU(2)$ doublets are $\tilde{h}_{1}=(\tilde{h}_{1L}^{0},\tilde{h}_{1L}^{-})^{T}$
and $\tilde{h}_{2}=(\tilde{h}_{2L}^{+},\tilde{h}_{2L}^{0})^{T}$ respectively.
Together with the two charged gauginos $\tilde{W}_{L}^{+}$ and $\tilde{W}_{L}^{-}$,
the charginos are combined into two left- and right-handed four component
spinors as $\psi_{L}=(\tilde{W}_{L}^{+},\tilde{h}_{2L}^{+})^{T}$
and $\psi_{R}=((\tilde{W}_{L}^{-})^{c},(\tilde{h}_{1L}^{-})^{c})^{T}$.
The chargino mass term has the form $\bar{\psi}_{R}M(z)\psi_{L}$
in the wino-higgsino space with mass matrix \begin{equation}
M(z)=\left(\begin{array}{cc}
M_{2} & gH_{2}(z)\\
gH_{1}(z) & \mu\end{array}\right)\end{equation}
where $M_{2}$ and $\mu$ are soft supersymmetry breaking parameters
containing CP phases and $H_{1}(z)$ and $H_{2}(z)$ the Higgs vacuum
expectation values (VEVs). The mass matrix is diagonalized by a bi-unitary
transformation $V^{\dagger}M^{}U$. The explicit form of the two rotation
matrices are \begin{eqnarray}
U & = & \frac{\sqrt{2}}{\sqrt{\Lambda(\Lambda+\Delta)}}\left(\begin{array}{cc}
\frac{1}{2}(\Lambda+\Delta) & -a\\
a^{*} & \frac{1}{2}(\Lambda+\Delta)\end{array}\right)\\
V & = & \frac{\sqrt{2}}{\sqrt{\Lambda(\Lambda+\bar{\Delta})}}\left(\begin{array}{cc}
\frac{1}{2}(\Lambda+\bar{\Delta}) & -\bar{a}\\
\bar{a}^{*} & \frac{1}{2}(\Lambda+\bar{\Delta})\end{array}\right)\end{eqnarray}
with \begin{eqnarray}
a=M_{2}u_{1}+\mu^{*}u_{2} & , & \bar{a}=M_{2}^{*}u_{2}+\mu u_{1}\\
\Delta=\left|M_{2}\right|^{2}-\left|\mu\right|^{2}+u_{2}^{2}-u_{1}^{2} & , & \bar{\Delta}=\left|M_{2}\right|^{2}-\left|\mu\right|^{2}-u_{2}^{2}+u_{1}^{2}\\
\Lambda=\sqrt{\Delta^{2}+4\left|a\right|^{2}} & , & u_{i}=gH_{i}(z)\end{eqnarray}
The two mass eigenvalues are given by $m_{1,2}=(\left|M_{2}\right|^{2}+\left|\mu\right|^{2}+u_{2}^{2}+u_{1}^{2}\pm\Lambda)/2$.
We are interested in the states which is Higgsino-like since only
Higgsino can be efficiently transported. The Higgsino-like state is
identified as particle 2(1) for parameter region $M_{2}>|\mu|(M_{2}<\left|\mu\right|)$.
From the above expression we obtain the off-diagonal elements of matrices
$\Sigma$ and $\Pi$ which is relevant to the mixing induced source.
\begin{eqnarray}
\Sigma_{12} & = & -\frac{1}{\Lambda(\Lambda+\Delta)}\left[(\Lambda+\Delta)a'-(\Lambda'+\Delta')a\right]\\
\Pi_{12} & = & -\frac{1}{\Lambda(\Lambda+\bar{\Delta})}\left[(\Lambda+\bar{\Delta})\bar{a}'-(\Lambda'+\bar{\Delta}')\bar{a}\right]\end{eqnarray}
the other off-diagonal elements $\Sigma_{21}(\Pi_{21})$ can be easily
obtained as $\Sigma(\Pi)$ is anti-Hermitian. The quantity $A_{11}$
relevant to the CP violating force term is given by \[
A_{11}=\frac{M_{2}\mu}{m_{1}^{2}\Lambda}(u_{1}u_{2})'\]

For simplicity, we only consider the chargino contribution as a dominant
source, as the transportation for neutrilinos is suppressed due to
much weaker couplings to fermions. The stop contributions are also
suppressed as they are bosons. Furthermore, the mass difference for
the two stops has to be very large. The left-handed stop $\tilde{t}_{L}$
must be above 1 TeV to evade the LEP experiment constraints \cite{Carena:2000id},
while the right-handed stop $\tilde{t}_{R}$ must be light a round
100 GeV in order to generate strongly first order phase transition
\cite{Carena1998,Laine1998}. The large mass difference between $\tilde{t}_{L}$
and $\tilde{t}_{R}$ leads to a very fast oscillation which strongly
suppressed the mixing-induced source in stop sector. As the semiclassical
force term does not present in the first order in derivative, we thus
neglect the stop contributions. 

\begin{figure}
\includegraphics[width=0.45\linewidth]{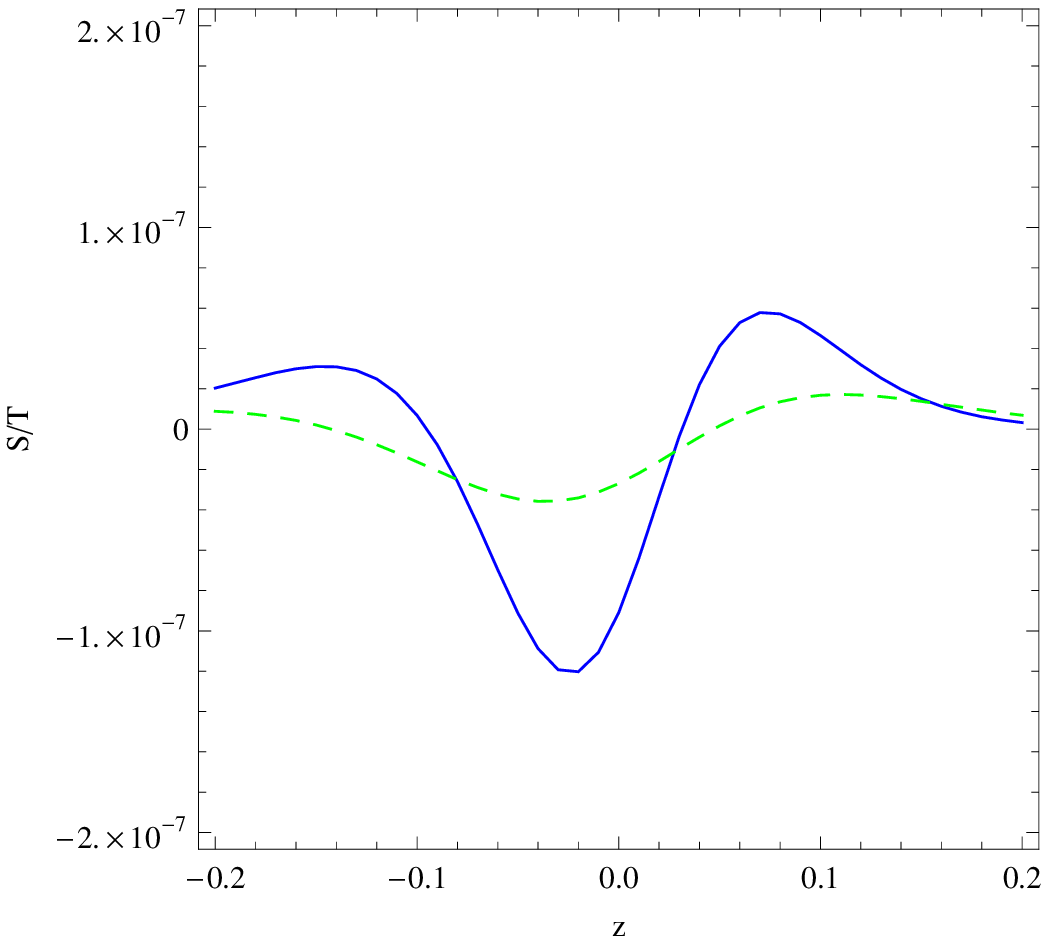}
\includegraphics[width=0.45\textwidth]{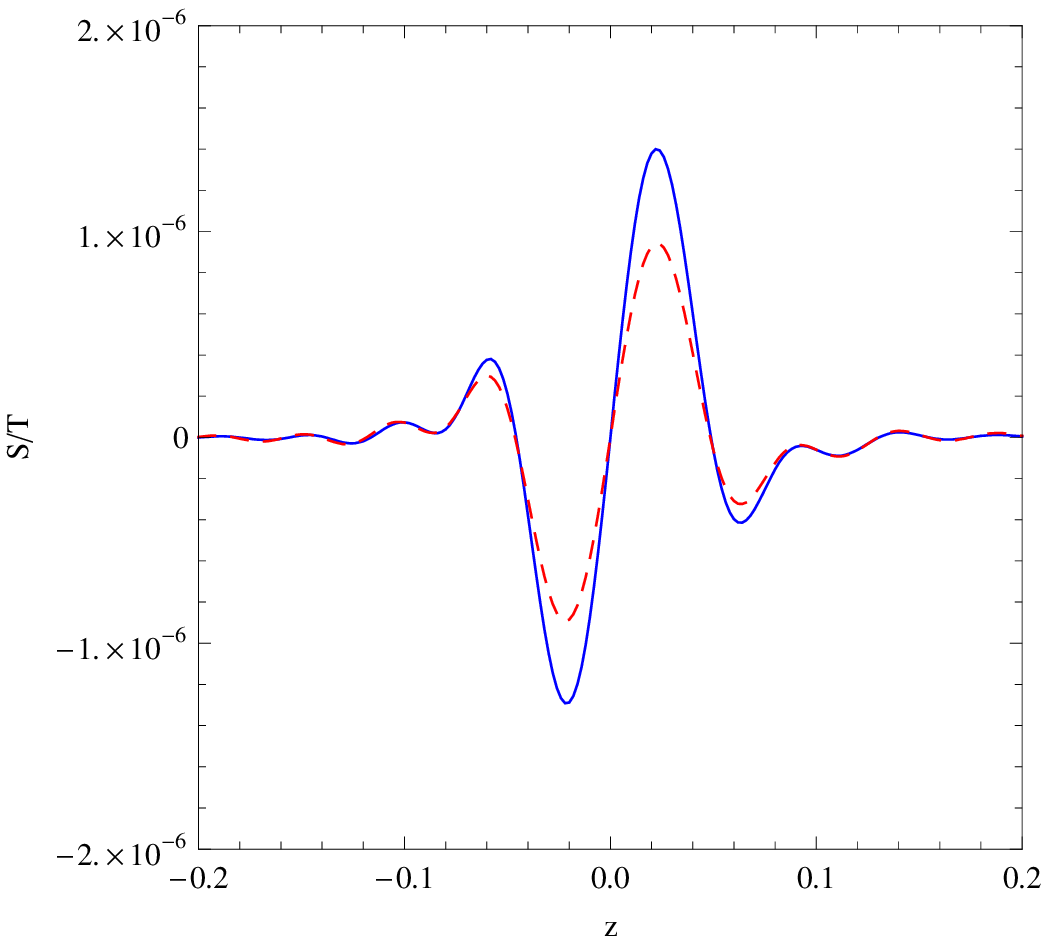}

\caption{(a)Left, the semiclassical force induced source term as a function
of $z$ ($\mbox{GeV}^{-1}$). (b) Right, the mixing-induced source
term as a function of $z$. The two curves in each plot corresponds
to $L_{w}=10/T$ (solid) and $15/T$ (dashed) respectively. The MSSM
parameters are fixed at $M_{2}=150$ GeV and $\mu=100$ GeV, with
$\phi_{\mu}=0.02$.}
\label{fig:1}
\end{figure}

For numerical illustrations we take the following simple wall profile\[
u(z)=g\sqrt{H_{1}^{2}(z)+H_{2}^{2}(z)}=g\frac{v_{c}}{2}\cdot\frac{1}{2}\left[1-\tanh\left(\frac{z}{L_{w}}\right)\right]\]
with $H_{1}(z)=u(z)\sin\beta$ and $H_{2}(z)=u(z)\cos\beta$ and $g$
the weak gauge coupling. $v_{c}$ is the Higgs VEV at the critical
temperature $T$ which is normalized at $v_{T=0}\simeq246$ GeV. For
numerical calculation we take the following reference values\[
v_{c}=120\mbox{GeV},T=90\mbox{GeV},\tan\beta=3,v_{w}=0.03\]
We consider different sets of wall-width and MSSM parameters which
is sensitive to the mixing-induced source. In Fig.\ref{fig:1}, we
give the source term rescaled by $1/T$ from both semiclassical force
and mixing-induced source term for wall width $L_{w}=10/T$ and $15/T$
respectively. The MSSM softbreaking parameters are fixed at $M_{2}=200$GeV
and $\left|\mu\right|=100$GeV corresponding to a chargino mass differences
$m_{1}^{}-m_{2}^{}=111$ GeV in the broken phase. The CP phase $\phi_{\mu}$
is set to a typical of $\phi_{\mu}=0.02$ which is a typically allowed
value after the constraints from the electron and Hg electric dipole
moments (EDMs) \cite{Lee2005,Regan:2002ta,Romalis:2000mg}. 
Note that the EDM constraints on CP phases depends on MSSM parameters
such as sfermion masses and CP-odd Higgs masses. For heavy left-handed
stop mass above a few TeV as also favored by the LEP limit, the
allowed CP phase can be much larger. Furthermore the two-loop
corrections to EDM are controlled by CP-odd Higgs mass, which can be
weakened by heavy Higgs mass, and has no significant effects on the source
term. Therefore, although small $\phi_{\mu}$ is used in the
calculation, the possibility of large CP phase close to the maximum
may still be allowed in this case.
The curves given in Fig.\ref{fig:1} show that both of source terms have
nontrivial spatial dependencies, but their origin are quite different.
The variation of the force term comes from the third derivative of
the kink-type Higgs condensate, which has one minimal and two maximums.
While the mixing-induced source term varies due to both the wall profile
variation and the oscillation. The oscillation leads to multiple local
minimum appearing in the curve and is suppressed at large distance
by the wall profile. For $M_{2}$ around 200 GeV, the mixing induced
source term peaks at $z\simeq0.03$ with an amplitude $S_{M}/T\simeq1.6\times10^{-6}$,
much larger than that from semiclassical force which peaks at $z\simeq0.08$
with $S_{F}/T=0.6\times10^{-7}$ %

\begin{figure}
\includegraphics[width=0.45\linewidth]{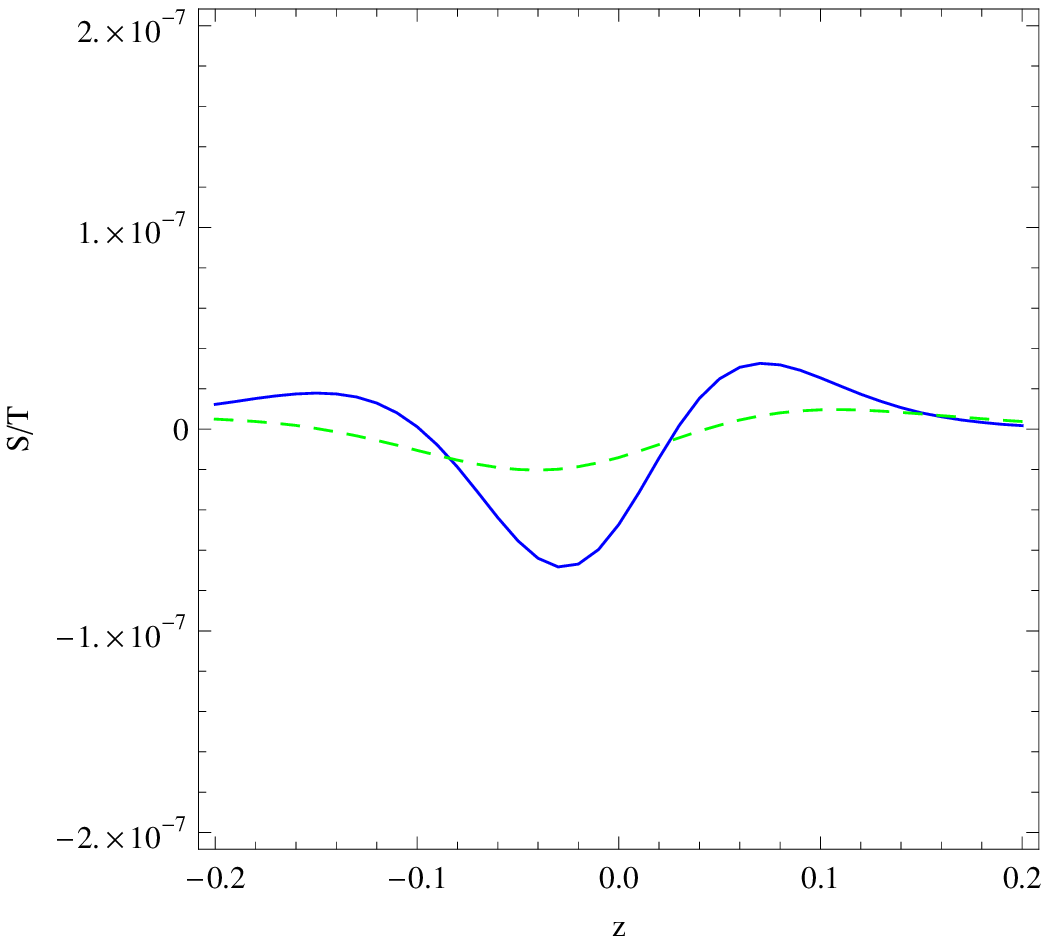}
\includegraphics[width=0.45\textwidth]{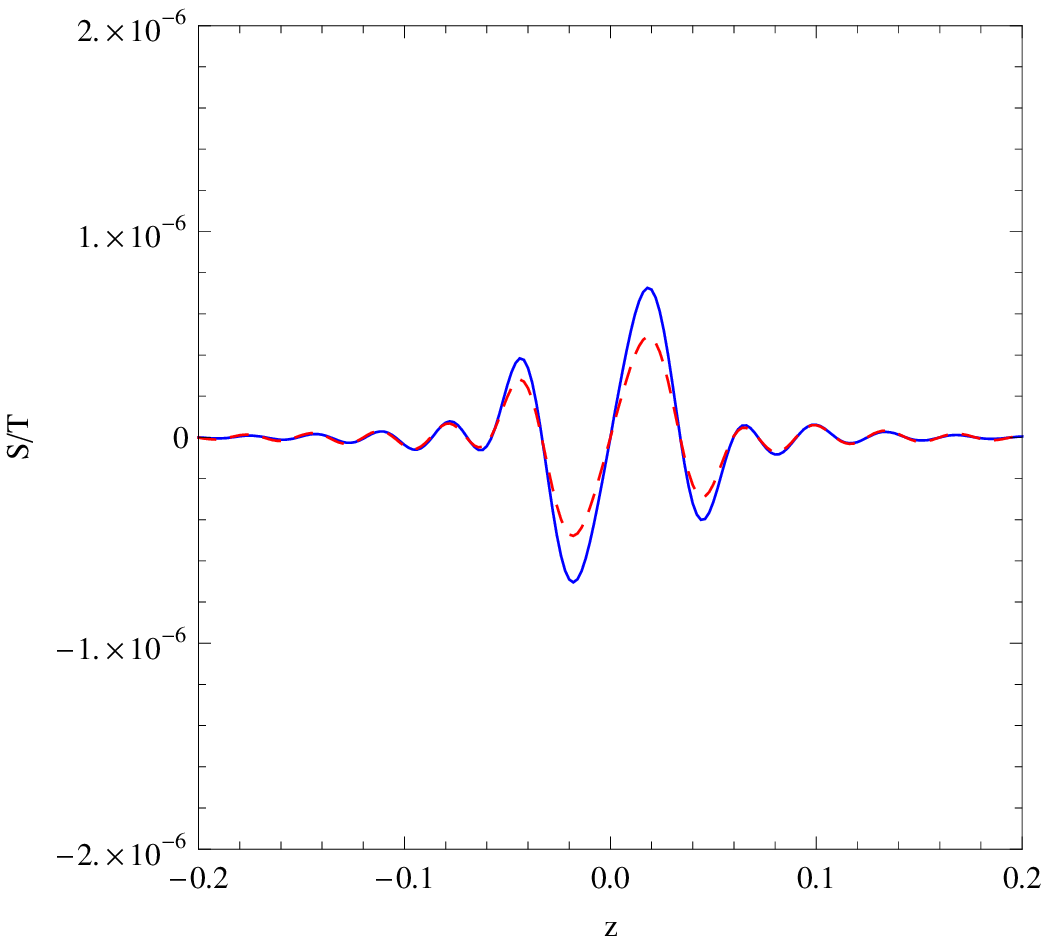}

\caption{(a)Left, the semiclassical force induced source term as a function
of $z$($\mbox{GeV}^{-1}$). (b) Right, the mixing-induced source
term as a function of $z$. The two curves in each plot corresponds
to $L_{w}=10/T$ (solid) and $15/T$ (dashed) respectively. The MSSM
parameters are fixed at $M_{2}=200$ GeV and $\mu=100$ GeV, with
$\phi_{\mu}=0.02$.}
\label{fig:2}
\end{figure}

In Fig.\ref{fig:2}, we give the results for a larger $M_{2}=250$
GeV, and still fix $\left|\mu\right|=100$GeV, corresponding to a
larger chargino mass differences $m_{1}-m_{2}^{}=157$ GeV in the
broken phase. One sees that for a larger chargino mass difference
both the source terms becomes smaller as they are $1/\Lambda$ suppressed.
The oscillation in the mixing-induced source term becomes obvious
and the wave length is shorter, thus the oscillation suppression is
stronger. The mixing induced source term peaks at $z\simeq0.03$ with
an amplitude $S_{M}/T\simeq8\times10^{-7}$. Although significantly
reduced, it still much larger than that from semiclassical force which
peaks at $z\simeq0.08$ with $S_{F}/T\simeq4\times10^{-8}$. 

\begin{table}
\input{table.tex}

\caption{Averaged source term (in unit of $10^{-8}$) for mixing induced source
and the force term. For wall width $L_{w}=10/T$ and $15/T$ respectively.}
\label{tab:averagedS}
\end{table}
To estimate the oscillation suppression effects it is useful to define
an averaged source over the wall width\begin{equation}
\bar{S}_{M(F)}\equiv\frac{1}{TL_{w}}\int_{0}^{L_{w}}S_{M(F)}(z)dz\end{equation}
We calculate the averaged source numerically for different $M_{2}=200\sim500$
GeV and list the results in Tab.\ref{tab:averagedS}. One sees that
$\bar{S}_{M}$ dominates over $\bar{S}_{F}$ in the range $200\mbox{GeV}\lesssim M_{2}\lesssim350\mbox{GeV}$.
With the value of $M_{2}$ increasing, the averaged source term $\bar{S}_{M}$
drops rapidly. For $L_{w}=10/T$, at $M_{2}=350$ GeV, the mixing
induced source is only $5.4\%$ of that at $M_{2}=200$ GeV, the suppression
is due to the increased oscillation frequency. The suppression in
force term $\bar{S}_{F}$ is mainly from the factor $1/\Lambda$,
which  decreases much slower. At $M_{2}=350$ GeV, it is still about
$30\%$ as large as that at $M_{2}=200$ GeV. At $M_{2}=200$ GeV,
their relative size is $\bar{S}_{M}/\bar{S}_{F}=14.6$. For large
$M_{2}=350$ GeV, although they are close in size, the mixing source
still dominates with $\bar{S}_{M}/\bar{S}_{F}=2.63$. This dominance
has a mild dependence on the wall width. For $L_{w}=15/T$, the relative
size between the two kind of sources remains roughly the same, although
both of the source term becomes smaller. The mixing-induced source
(semiclassical force) term is $\sim40\%(\sim30\%)$ of that at $L_{w}=10/T$.
When the value of $M_{2}$ is around 450 GeV, the two type of source
term becomes comparable in size. For a very large $M_{2}=500$ GeV,
the semiclassical force term becomes dominate, and  the mixing-induced
source term is about an order of magnitude smaller. 

In conclusion, we have studied the effects of flavor mixing in a generalized
WKB approach in which the off-diagonal terms in the equation of motion
are taken into account as perturbations. With the presence of a slowly
moving CP violating bubble wall, an extra mixing-induced CP violating
source appears which exhibit an oscillation behavior in analogy to
the neutrino mixings. In two flavor mixing case, the oscillation frequency
is proportional to the difference of mass square. The size of the
mixing-induced source could be larger than that from the conventional
semiclassical force as it is at the first order in derivative expansion.
We have made a numerical study of the oscillation suppression effects
for chargino case in MSSM. For a small $\left|\mu\right|=100$ GeV,
in a large range $200\lesssim M_{2}\lesssim350$ GeV, the mixing -induced
source term dominates over the semiclassical force term. The method
is valid for small mixing case, a more general method is needed to
deal with the maximum mixing case. However, even in the small mixing
case the mixing-induced source already indicate that a significant
enhancement of the final baryon number asymmetry is possible, which
will relax the tension between the observed baryon number asymmetry
in the universe and the constraints from electron and atom EDM in
the MSSM.

\begin{acknowledgments}
The author is grateful to Y. Okada, E. Senaha, M. Asano and
S. Kanemura for helpful discussions. The work at KITP was supported in
part by the National Science Foundation under Grant No. PHY05-51164

\bibliographystyle{aalpha}
\bibliography{ewvg2}

\end{acknowledgments}

\end{document}

%% file: table.tex
\begin{tabular}{ccccccc}
\hline\hline
 $M_2(GeV)$      	&200  	&250 	&300 	& 350 	&450 	&500\\
\hline
$\bar{S}_M(L_w=10/T)$  	&27.5 	&8.29 	&3.37   &1.50 	&0.243	&0.0364\\
$\bar{S}_F(L_w=10/T)$  	&1.89 	&1.23	&0.822 	&0.575 	&0.317	&0.246\\
\hline
$\bar{S}_M(L_w=15/T)$  	&10.9 	&5.48 	&1.45 	&0.604  &0.121	&0.0177\\ 
$\bar{S}_F(L_w=15/T)$  	&0.561	&0.365	&0.243	&0.170 	&0.094 	&0.073\\
\hline\hline
\end{tabular}